\documentclass[a4paper,11pt]{article}
\usepackage{pos} 
\usepackage{graphicx}
\usepackage{epsfig}
\usepackage{bm}
\usepackage{units}
\usepackage{upgreek}
\usepackage{latexsym,amsmath,wasysym,float}
\usepackage{color}
\usepackage{scrextend}
\usepackage{mathrsfs}
\usepackage{enumitem}
\usepackage[export]{adjustbox}
\usepackage{caption}

\title{Prospects for PBR detection of KM3-230213A-like events}

\author*[a]{Angela V. Olinto}
\author[b]{Luis A. Anchordoqui}
\author[c]{Austin Cummings}
\author[a]{Johannes~Eser}
\author[d]{Diksha Garg}
\author[e]{Claire Gu\'epin}
\author[f]{Tobias Heibges}
\author[g]{John F. Krizmanic}
\author[a]{Thomas~C.~Paul}
\author[b]{Karem Pe\~nal\'o Castillo}
\author[d]{Mary Hall Reno}
\author[g]{Tonia~M.~Venters}

\affiliation[a]{Columbia Astrophysics Laboratory, Columbia University, 538 West 120th Street , New York, NY, USA}
\affiliation[b]{Lehman College, City University of New York, NY 10468, USA}
\affiliation[c]{Pennsylvania State University, PA 16802, USA}
\affiliation[d]{University of Iowa, IA 52242, USA}
\affiliation[e]{Laboratoire Univers et Particules de Montpellier, Montpellier, France}
\affiliation[f]{Department of Physics, Colorado School of Mines,  CO 80401, USA}
\affiliation[g]{NASA Goddard Space Flight Center, MD 20771, USA}

\onbehalf{for the JEM-EUSO collaboration} 

\emailAdd{avo2113@columbia.edu}

\abstract{POEMMA-Balloon with Radio (PBR) is a scaled-down version of the Probe
Of Extreme Multi-Messenger Astrophysics (POEMMA) design, optimized to
be flown as a payload on one of NASA's sub-orbital super pressure
balloons circling the Earth above the southern oceans for a mission duration of
more than 20 days. One of the main science objectives of PBR is to
follow up astrophysical event alerts in search of neutrinos with very
high energy ($10^8 \lesssim E_\nu /{\rm GeV} \lesssim 10^{10}$). Of particular
interest for anticipated PBR observations, the KM3NeT Collaboration has recently reported the detection of the neutrino KM3-230213A with $10^{7.8} \lesssim E_\nu/{\rm GeV} \lesssim 10^{9.0}$. Such an unprecedented event is in tension with upper limits on the cosmic neutrino flux from IceCube and the Pierre
Auger Observatory: for a diffuse isotropic neutrino flux there is a
$3.5\sigma$ tension between KM3NeT and IceCube measurements, and about $2.6\sigma$ if the neutrino flux originates
in transient sources. Therefore, if
KM3-230213A was not beginner's
luck, it becomes compelling to consider  beyond Standard Model (BSM)
possibilities which could lead to a signal at KM3NeT-ARCA but not at
IceCube/Auger. We calculate the PBR horizon-range sensitivity to probe
BSM physics compatible with observation at KM3NeT-ARCA and
non-observation at 
IceCube/Auger. As an illustration, we consider a particular class of BSM physics models which
has been described in the literature as a possible explanation of KM3-230213A.}

\ConferenceLogo{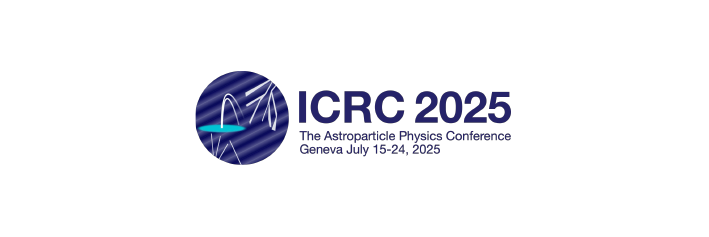}

\FullConference{39th International Cosmic Ray Conference (ICRC2025)\\
 15–24 July 2025\\
Geneva, Switzerland\\}

\begin{document}
\maketitle

\FullConference{39th International Cosmic Ray Conference (ICRC2025)\\
 15–24 July 2025\\
Geneva, Switzerland\\}

\section{Introduction}
\label{sec:intro}

On February 13, 2023, a partial implementation of the KM3NeT/ARCA
detector~\cite{KM3Net:2016zxf, KM3NeT:2022pnv} recorded an energetic,
nearly horizontal muon track, with a reconstructed energy of
$ E_{\mu}=\unit[120^{+110}_{-60}]{PeV}$ \cite{KM3NeT:2025npi}. This
event, designated as KM3-230213A, most likely was initiated by a neutrino interaction. It represents the highest-energy
neutrino event observed to date. To optimally study the possible
origin of such an extreme energy neutrino, an instrument is desired
that is capable of pointing at any position in the sky, following the
source for a sufficient time, and having a low energy threshold.

POEMMA-Balloon with Radio (PBR) is a stratospheric balloon mission~\cite{ICRC2025Eser} representing the most advanced pathfinder for
future space-based missions, such as the Probe of Multi-Messenger
Astrophysics (POEMMA)~\cite{POEMMA:2020ykm}. PBR is designed to
fulfill these above requirements as one of its three main scientific
objectives: the search for transient astrophysical neutrinos via a
Target-of-Opportunity (ToO) approach following multi-messenger event
alerts. In addition, its scientific program includes the observation
of high-altitude horizontal air-showers  and the inaugural 
measurement of the fluorescence emissions from extensive air showers
(EASs) generated by ultra-high-energy cosmic rays (UHECRs) from a
suborbital platform. PBR is scheduled to fly as a payload on a NASA Super Pressure Balloon (SPB), which will circumnavigate the Southern Ocean for more than 20 days following its launch from Wanaka, New Zealand, in Spring 2027.

As a pathfinder, PBR builds on the design studies of POEMMA and
leverages experience from previous balloon missions, including
EUSO-SPB1~\cite{JEM-EUSO:2023ypf} and EUSO-SPB2~\cite{Adams:2025owi}. The payload integrates three primary instruments: a Fluorescence Camera (FC), a Cherenkov Camera (CC), and a dedicated Radio Instrument (RI). The FC, which records the longitudinal profiles of UHECR-induced air showers, is described in detail elsewhere~\cite{Battisti:2024rfv}. The CC is incorporated into the hybrid focal surface of a \unit[1.1]{m} Schmidt telescope with
a field of view (FoV) of 36$^{\circ}$ (vertical) by 24$^{\circ}$
(horizontal). Together with the RI, which complements the optical
measurements, these instruments form the principal detectors for
neutrino signatures such as those associated with event KM3-230213A. A
NASA-supplied rotator paired with a custom-designed tilt system
facilitates precise positioning of both instruments to any location
below the horizon.

The CC is designed to capture ultra-fast optical pulses,
operating with an integration time of only \unit[10]{ns}. It uses
silicon photomultipliers with a pixel size of
$3~{\rm mm} \times 3~{\rm mm}$, yielding an instantaneous FoV of
$0.2^\circ$ per pixel. The 2048-pixel camera, with a detection
wavelength range from  320--900~nm provides a total FoV of $6^\circ$ (vertical) by $12^\circ$ (horizontal). This configuration is sufficiently wide to monitor an astrophysical event for over 20 minutes without active tracking. The estimated energy threshold is approximately \unit[500]{TeV}.

The RI consists of two sinusoidal radio antennas with a broadband gain between 50--500\,MHz, based on the design of the Low Frequency instrument for the Payload for Ultra-High Energy (PUEO) experiment and optimized for EAS detection \cite{Abarr_2021}. Mounted beneath the telescope, these antennas cover the entire optical FoV with 5dB points of $\pm30^{\circ}$ from boresight, and are externally triggered by the CC.

In this paper we
calculate the PBR sensitivity to KM3-230213A-like
events. The layout is as follows. In Sec.~\ref{sec:1} we rexamine
the main characteristics of the KM3-230213A event, in Sec.~\ref{sec:3}
we explore potential interpretations of KM3-230213A that can be
investigated with PBR, in Sec.~\ref{sec:4} we evaluate the sensitivity
of PBR to such phenomena, and in Sec.~\ref{sec:5} we draw our conclusions.

\section{The KM3-230213A  event}
\label{sec:1}

\begin{figure}
    \centering \includegraphics[width=0.5\linewidth,valign=c]{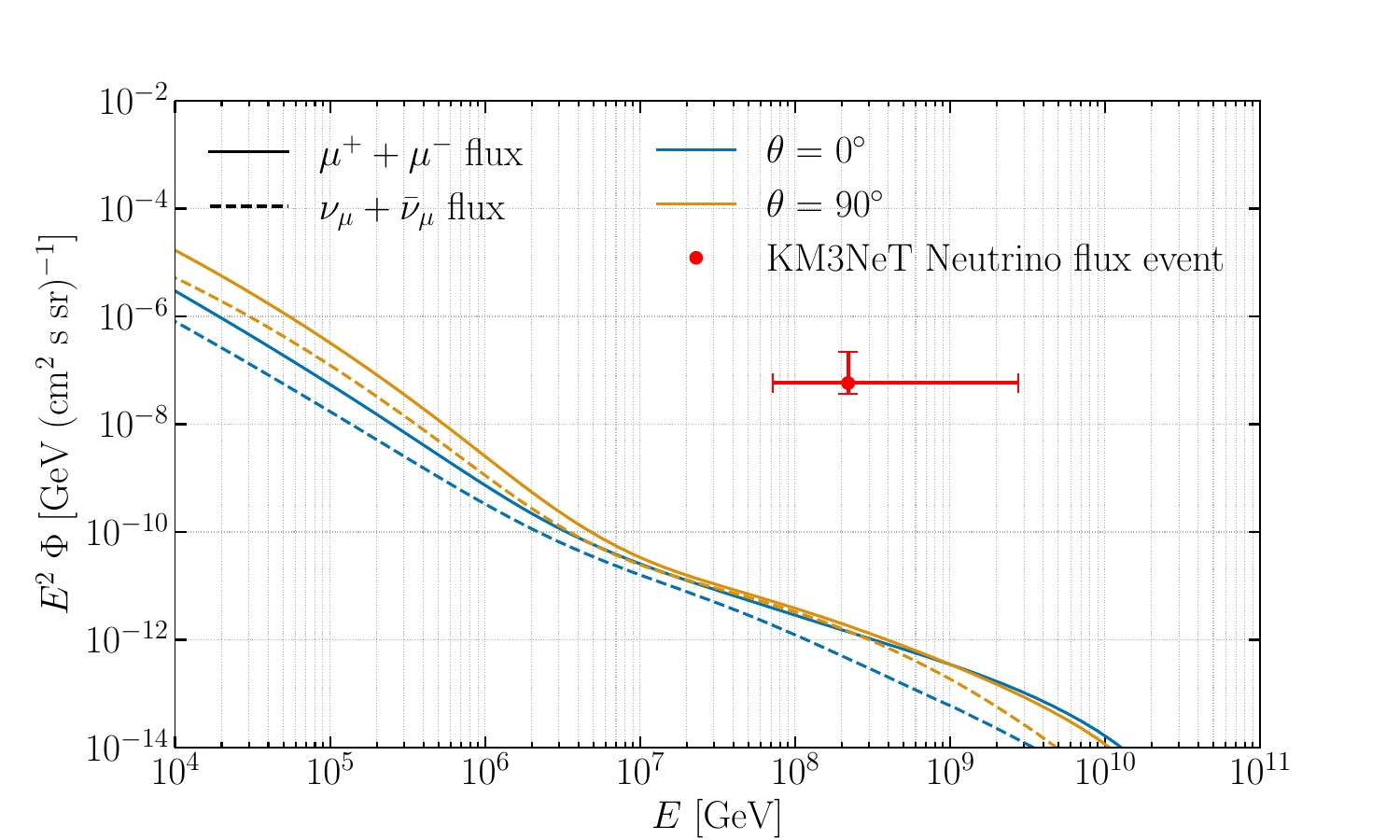}
    \hfil
    \begin{minipage}{\dimexpr 0.5\linewidth-\columnsep}
        \captionsetup{singlelinecheck=off, skip=0pt}
      \caption{Flux of atmospheric muons scaled by muon energy $E_\mu^2$ (solid lines) and atmospheric muon neutrinos scaled by neutrino energy (dashed lines) for zenith angles $0^\circ$ and $90^\circ$  evaluated using MCEq \cite{Fedynitch:2018cbl}
    with Sibyll 2.3c and the H3p cosmic ray flux, and the KM3NeT scaled isotropic per flavor neutrino flux (data point with error bars) \cite{KM3NeT:2025npi}.}
    \label{fig:atm-muons} 
    \end{minipage}
    \vspace{-2ex}
\end{figure}

The highest energy muon event observed by KM3NeT/ARCA during a 284.7
day observation period (using 21 detection strings) recorded muon
energy of $120^{+110}_{-60}$ PeV, with a 90\% CL interval of 35--380
PeV \cite{KM3NeT:2025npi}. The muon's trajectory was reconstructed to
$0.6^\circ$ above the horizon and at $259.8^\circ$ in azimuth, with a
directional uncertainty is $1.5^\circ$ (68\% CL).  This uncertainty is
anticipated to improve to $0.12^\circ$ once detector positioning is
more precicely determined.  The event is labeled KM3-230213A, with
event coordinate central values of ${\rm RA}=94.3^\circ$ and ${\rm
  DEC}= -7.8^\circ$, and MJD$= 59988.0533299$.
If the event comes from a steady isotropic neutrino flux, its scaled per flavor flux is $E_\nu^2\Phi(E_\nu)=
5.8^{+10.1}_{-3.7}\times 10^{-8}$ GeV/(cm$^2$s\,sr) \cite{KM3NeT:2025npi}, shown by the red data point in Fig.~\ref{fig:atm-muons}.

The KM3NeT collaboration has ruled out atmospheric muons as the origin
of the event.  Their analysis used MCEq software \cite{Fedynitch:2018cbl} to model
atmospheric particle cascades and calculate sea-level muon fluxes,
employing the SIBYLL 2.3c hadronic interaction model.
The solid curves in Fig.~\ref{fig:atm-muons} show
the sea-level flux of atmospheric muons scaled by $E_\mu^2$ evaluated using MCEq with the H3p cosmic ray spectrum and composition
\cite{Gaisser:2011klf}.
For reference, the dashed lines show the atmospheric flux of $\nu_\mu+\bar{\nu}_\mu$.
The lower and upper pairs of curves are for vertical ($\theta = 0^\circ$) and
horizontal ($\theta=90^\circ$) directions, respectively. Above muon energies of
$\sim 10$ PeV, the atmospheric muon flux is nearly isotropic since
the principal contributions come from prompt decays of charm hadrons and
light unflavored mesons \cite{Illana:2009qv,Illana:2010gh}.

Above $\sim 10$ PeV, the atmospheric muon flux at sea level predicted
by MCEq scales as $\sim E^{-2.9}$, yielding an integrated flux of
$\sim 1.4\times 10^{-19}$ muons/(cm$^2$s\,sr) above 35 PeV. In
Fig.~\ref{fig:atm-muons}, the atmospheric muon flux above
10 PeV is $E_\mu^2\phi_\mu \sim 3.6\times 10^{-12}$(100 PeV/$E_\mu)^{0.9}$\,GeV/(cm$^2$s\,sr).
This corresponds to $\sim 3.5\times 10^{-12}$ muons/(cm$^2$sr) at sea level for the 287.4 days
of data-taking (100\%\ duty cycle). For a cone apex angle of $3^\circ = 2\cdot 1.5^\circ$ and
detector area of $\sim 0.4$ km$^2$, the number of muons with energies
greater than 35 PeV at sea level that could potentially contribute
directly to the KM3NeT event is $\sim 2\times 10^{-3}$.  For a muon
energy of 120 PeV, at sea level, the number of muons is approximately
one order of magnitude lower than this.

Muons at sea level lose energy in transit to
the detector. The KM3NeT collaboration reports a 300 kmwe column depth from sea level to detector for
the most likely track trajectory direction of $0.6^\circ$ above the horizon. They report that the trajectory column depth from sea level to the detector would be $\sim 60$ kmwe
(all water) if its direction is $2.6^\circ$ above the horizon. A
conservative estimate of muon energy loss from the average energy loss
formula, $\langle dE_\mu/dX\rangle \simeq - b E_\mu$,
can be obtained using $b=3.5\times 10^{-6}$cm$^2$/g
\cite{Groom:2001kq,mu-eloss-web}. For the shorter column depth of $X= 60$
kmwe, $E_\mu^f\simeq E_\mu^i\exp
(-bX)=35$ PeV requires an initial energy of $E_\mu^i\simeq 4.6\times 10^{16}$ GeV ($4.6\times 10^{10}$ PeV). 

\begin{figure*}
    \centering
    \includegraphics[width=0.45\linewidth]{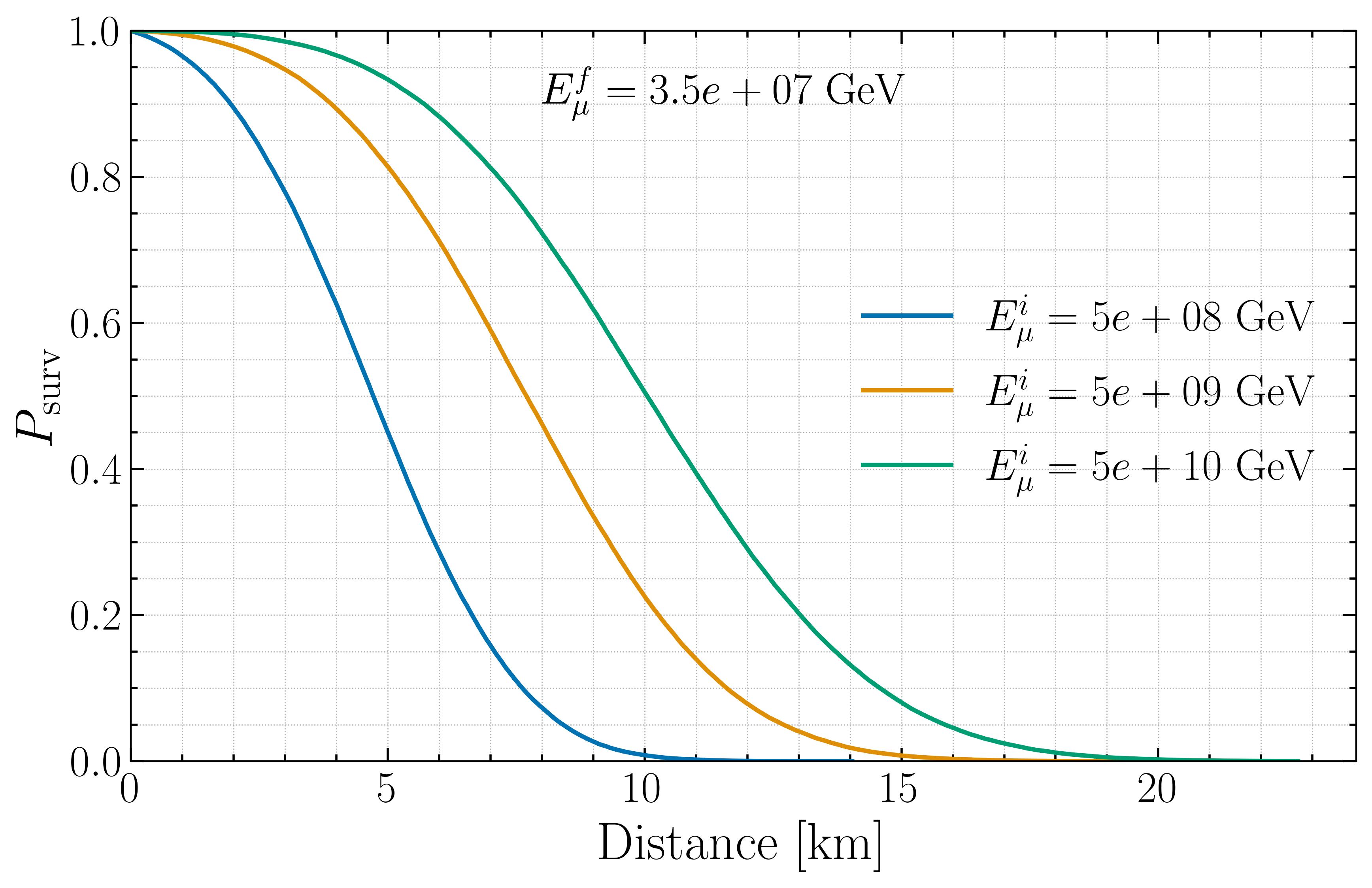}
    \includegraphics[width=0.46\linewidth]{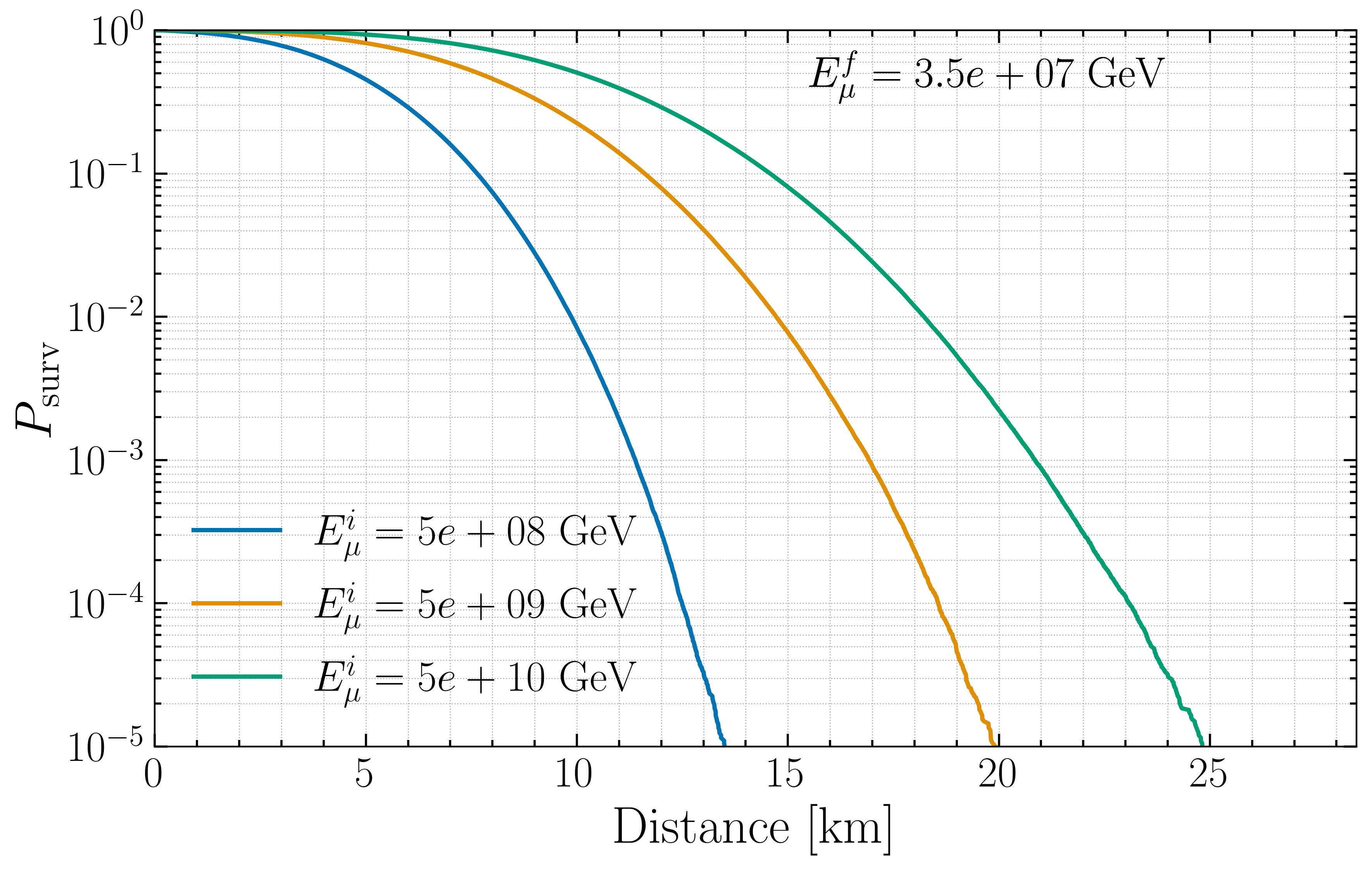}
    \caption{Survival probability of muons as a function of distance traveled in water ($\rho=1$ g/cm$^3$), for three initial energies with a final muon energy of $E_\mu^f = 35$ PeV. The right plot shows the same results with log-linear scales.}
    \label{fig:psurv-muons}
    \vspace{-2ex}
\end{figure*}

Muon electromagnetic
energy loss can be understood more precisely by including stochastic effects.
Using the ALLM photonuclear energy loss formula~\cite{Abramowicz:1997ms,Abramowicz:1991xz} along with bremsstrahlung and pair production processes \cite{Lohmann:1985qg}, one can find
the muon survival probability \cite{Lipari:1991ut,Dutta:2000hh}
as a function of distance in water. This is plotted in the left panel of 
Fig.~\ref{fig:psurv-muons} for three initial muon energies and a final muon energy greater than or equal to 35 PeV.
The right panel shows the same muon survival probability on a log scale to illustrate the
rapid fall-off of the survival probability as a function of distance for initial muon energies in the range of $5\times 10^8-5\times 10^{10}$ GeV and $E_\mu^f=35$ PeV. Even for an initial muon energy of $5\times 10^{10}$ GeV and $E_\mu^f=35$ PeV,
the survival probability is less than $10^{-5}$ for $X\simeq 25$ kmwe, and
negligible for 60 kmwe. The probability for a lower energy muon to
survive a distance of 60 kmwe with an energy of $E_\mu^f=35$~PeV is significantly smaller.
As noted by KM3NeT
Collaboration, even if the candidate muon energy is at the lower
end of the 90\%\ confidence limit of the energy measurement, the atmospheric
muon flux is not a candidate source of KM3-230213A.

\section{Possible neutrino origins of KM3-230213A}
\label{sec:3}

Having ruled out the possibility that KM3-230213A was engendered by an atmospheric
muon, the next step is to investigate the possible sources of cosmic
neutrinos. Actually, a neutrino origin of KM3-230213A is also
challenging because the Pierre Auger Observatory and IceCube, which
have been operating with a much larger effective area than KM3NeT and
for a considerably longer time, have not observed any neutrinos with
$E_\nu \gtrsim 10^{7.6}~{\rm
  GeV}$~\cite{PierreAuger:2023pjg,IceCube:2025ezc}. The absence of
extreme energy neutrinos in the Auger~\cite{PierreAuger:2023pjg} or IceCube~\cite{IceCube:2025ezc} data samples excludes any association
of KM3-230213A with the diffuse cosmogenic neutrino flux. To be more specific, for a diffuse isotropic neutrino flux there is a
$3.5\sigma$ tension between KM3NeT-ARCA and IceCube measurements. The tension between KM3NeT-ARCA and IceCube measurements is about $2.6\sigma$ if the neutrino flux originates
in a transient source~\cite{Li:2025tqf,KM3NeT:2025ccp}. The combined observations from KM3NeT-ARCA,
IceCube, and Auger are consistent with a
source flare of duration $T \sim 1~{\rm yr}$, with muon neutrino flux
$F \sim 2 \times 10^{-7} \ (1~{\rm yr}/T)~{\rm
  GeV/(cm^2s)}$~\cite{Neronov:2025jfj}. The all-sky rate of similar neutrino flaring sources is
constrained to be ${\cal R} \lesssim 0.4/{\rm yr}$. The population of transient sources proposed in~\cite{Neronov:2025jfj} has a cosmological distance scale of 20~Gpc. Similar results
have been reported in~\cite{Yuan:2025zwe}. 

To elucidate the origin of KM3-230213A it is pivotal to examine
consistency with multi-messenger observations; particularly,
constraints from $\gamma$-ray telescopes. It is of common knowledge that transparent astrophysical sources of
extreme energy neutrinos would also produce a bright $\gamma$-ray
signal. To date, no convincing evidence of $\gamma$-rays
accompanying KM3-230213A has been observed~\cite{Airoldi:2025opo}. As a consequence, constraints can be placed on a combination of
the source redshift and the intergalactic magnetic field strength
between the source and
Earth~\cite{Fang:2025nzg,Crnogorcevic:2025vou}. If the strength of the
intergalactic magnetic field were $B \gtrsim 3 \times 10^{-13}~{\rm G}$,
then the $\gamma$-ray emission from farway sources could be attenuated {\it en route}
to Earth, and the model of transient sources described
in~\cite{Neronov:2025jfj}  would be consistent with observations. On the other hand, the absence of a $\gamma$-ray signal associated to KM3-230213A
challenges scenarios in which this neutrino has a Galactic origin,
e.g., from the decay of super-heavy dark matter clustered in the halo. Nevertheless, it was
pointed out in~\cite{Klipfel:2025jql} that if the dark matter were just
ordinary matter locked inside primordial black holes (PBHs), then the KM3NeT-ARCA event might have originated from a Galactic transient
explosion of a PBH near the end of its evaporation
lifetime. Actually, to accommodate the $\gamma$-ray
constraints one should advocate for scenarios with
large extra dimensions, because 
PBHs living in the higher-dimensional bulk would radiate SM singlets
that eventually oscillate into
active neutrinos~\cite{Anchordoqui:2025xug,Anchordoqui:2025opy}. It was recently noted that a cosmological population of microscopic black holes, which rapidly evaporate via Hawking radiation, emitting intense, short-lived bursts of neutrinos with $E_\nu \gtrsim 10^8{\rm GeV}$ can accommodate the groundbreaking KM3NeT-ARCA detection while remaining consistent with IceCube’s non-detections~\cite{Sakharov:2025oev}. These microscopic black holes are expected to be produced via bubble collisions generated in the merger of astrophysical black holes; therefore, exploring another window of multi-messenger astrophysics driven by neutrinos and gravitational waves. 

A plethora of models advocating beyond Standard Model (BSM) physics
have been proposed to explain the origin of KM3-230213A (we
refrain from referring to this literature in detail in these
proceedings). Of particular interest for PBR observations are BSM
possibilities which could lead to a signal at KM3NeT-ARCA but not at
IceCube/Auger. For example, the BSM model conjectured in~\cite{Brdar:2025azm} 
relies on the fact that the KM3-230213A
 has traversed approximately 147~km of rock and sea en route to the
 detector, whereas neutrinos arriving from the same location in the
 sky would have only traveled through about 14~km of ice before
 reaching IceCube. This is because the arrival direction of
 KM3-230213A has an incidence angle of $0.6^\circ$ at KM3NeT-ARCA and
 about $8^\circ$ above the horizon at IceCube. The idea proposed in~\cite{Brdar:2025azm} 
 is that a new physics matter potential can resonantly amplify sterile-to-active transitions and
 accommodate oscillations of extreme energy sterile neutrinos in the 100~km
 range.  The difference in
 propagation distance can then be used to increase the active neutrino
 flux near the KM3NeT detector, alleviating the tension between
 KM3NeT-ARCA and IceCube observations. This BSM model then requires that most of the emission by a
point source in the nominal direction of the detected KM3-230213A
neutrino arrives at Earth as sterile neutrinos. Two other similar
scenarios which make use of the difference in propagation distance are
described in~\cite{Farzan:2025ydi,Dev:2025czz}. PBR will sample a range of propagation distances. As a point source moves through the FoV of PBR's CC,  particle trajectories through the Earth range up to chord lengths of 2,300 km and column depths up to 7,400 kmwe (see, e.g., Fig.~2 in \cite{Reno:2019jtr}).

\section{PBR discovery reach}
\label{sec:4}

The integrated exposure of PBR is limited by observing conditions
(e.g., dark skies, presence of clouds), the length of time a given
source can be observed, and the overall duration of the balloon
flight. Nevertheless, the expected instantaneous
acceptance for very-high-energy neutrinos is expected to be
comparable with other ground based detectors. Thus, searches of very-high-energy neutrinos with PBR will be especially powerful for transient sources. 

To assess PBR's performance in observing KM3-230213A-like events,
we estimate its sensitivity to neutrino bursts. To that end, we calculate the time-averaged $\nu_{\tau}$ acceptance:
\begin{equation}
\langle A(E_{\nu},\phi,\delta)\rangle_{T_0} = \frac{1}{T_0}\int _{t_0}^{t_0+T_0}\, dt \ f(t)\, A(\beta_{\rm tr}(t,\lambda_{\oplus},\phi_{\oplus}),E_{\nu},\phi,\delta)
\ ,
\label{eq:avga}
\end{equation}
where $\phi$ and $\delta$ are the co-latitude and longitude of the celestial position of a given source (\textit{e.g.}, $\phi$ is the right ascension in the equatorial celestial coordinate system and $\delta$ is the declination), $T_0$ is the observation time, and $f(t)$ is an observation efficiency as a function of time that accounts for the reduction in the observation time due to light from the Sun and Moon~\cite{Venters:2019xwi}. In eq. (\ref{eq:avga}), $A(\beta_{\rm tr}(t,\lambda_{\oplus},\phi_{\oplus}),E_{\nu},\phi,\delta)$ is the instantaneous acceptance to $\nu_{\tau}$s from the source as defined by the geometry of the region of interest around the source on the surface of the Earth and the probability of detecting EASs induced by $\tau$ leptons that from the Earth at a selected location within the region (at \textit{e.g.,} Earth latitude, $\lambda_{\oplus}$, and longitude, $\phi_{\oplus}$) and that propagate along quasi-parallel trajectories that point back to the source. The trajectory of the $\tau$ lepton makes an angle $\beta_{\rm tr}$ above the Earth's surface. We perform Monte Carlo simulations to sample the region of interest and to calculate the detection probabilities of sampled events based on $\nu_{\tau}$ interaction physics in the Earth, $\tau$-lepton decays, EAS development, and the propagation of Cherenkov light signals through the atmosphere to the detector. In calculating time-averaged acceptances, we account for the changes in the source's position with respect to the Earth and the detector. We also account for model the balloon's trajectory based on historical wind patterns and assuming a launch on April 6, 2027. 

For the calculations we present here, we consider two scenarios: a
short, 1000-s burst of a transient source in an optimal observation
time window ($f(t)=1$), and a period of neutrino emission lasting
longer than $\sim 20$ days (as motivated by typical balloon mission
durations). In the short burst scenario, we assume the telescope is optimally pointed to catch the source just as it dips below the horizon at the burst's initial time $t_0$.

To determine the PBR discovery reach for neutrino sources, we adopt the
quasi-differential neutrino flux over a decade of
energy~\cite{Anchordoqui:2002vb}. The $\nu_\tau$--sensitivity can be approximated by the relation 
\begin{equation}\label{eq:sensitivity}
\nu_\tau{\rm -sensitivity} = \frac{2.44}{\ln 10}
\times\frac{E_{\nu}}{ \langle A(E_{\nu})\rangle_{T_0}} \ ,
\end{equation}
where we have
taken the $90\%$ CL assuming no signal and no background~\cite{Feldman:1997qc}, {\it viz.} $2.44/\ln 10$.

\begin{figure}[t] 
    \centering
    \begin{minipage}[t]{0.99\linewidth}
    \centering
\includegraphics[width=0.45\linewidth, trim = 5mm 15mm 90mm 30mm,
clip]{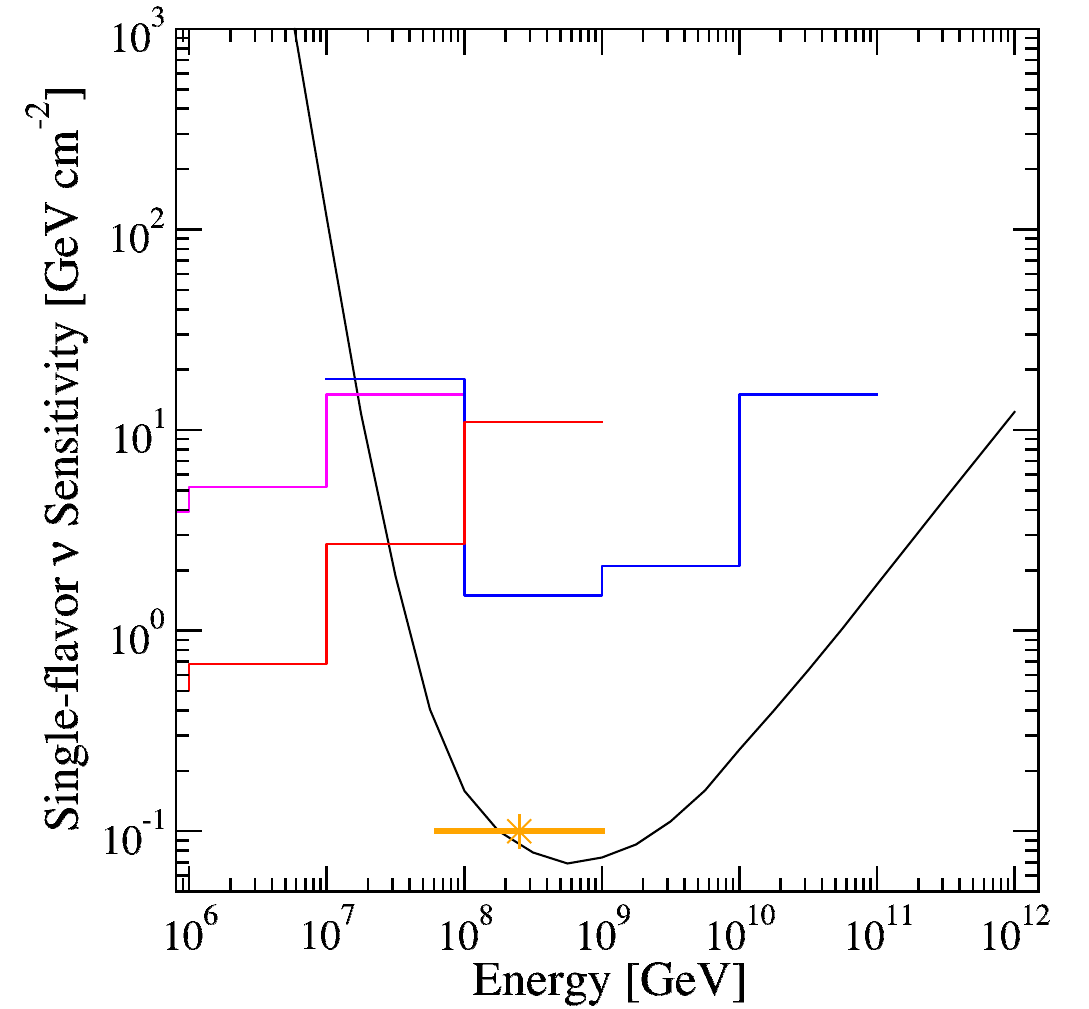}\includegraphics[width=0.45\linewidth,
trim = 5mm 15mm 90mm 30mm, clip]{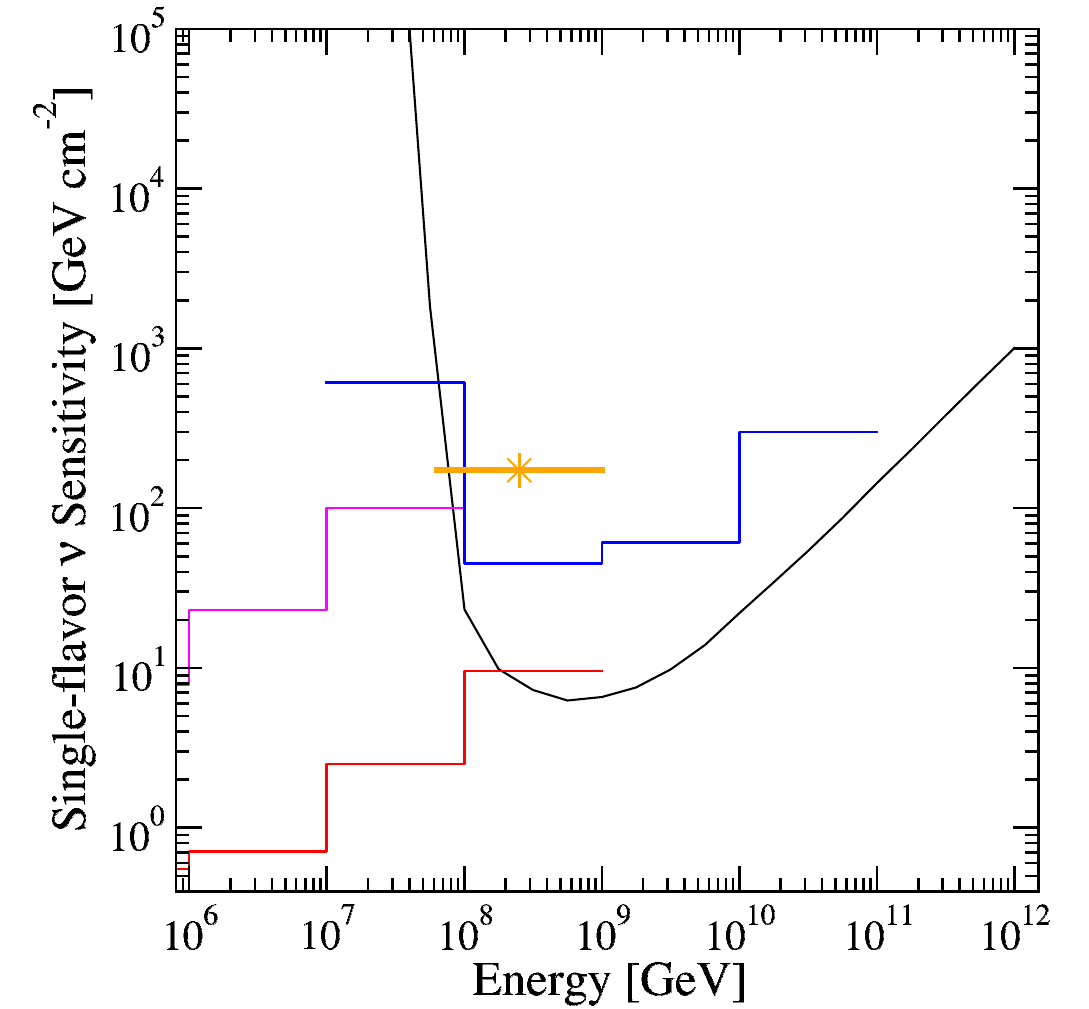}
        \caption{Single-flavor 90\%~CL $\nu_\tau$ sensitivity of PBR
          for optimal observing conditions during a 1000-s ToO
          observation (left) and over a 20-day period limited to
          dark-sky observations (right). For comparison, the modeled
          1000-s and 14-day fluences for a burst of BSM messengers
          (which produce $\tau$-leptons inside the Earth) are indicated by orange lines and stars. Also plotted are the upper limits for GW170817 reported by IceCube (red), Auger (blue), and ANTARES (magenta) computed for 1000 s and 14 days
        (see Ref.~\cite{ANTARES:2017bia}). \label{fig:3}} 
    \end{minipage}%
    \hfill
    \vspace{-2ex}
    \end{figure}

    \begin{figure}
    \centering
    \includegraphics[width=0.5\linewidth,valign=c]{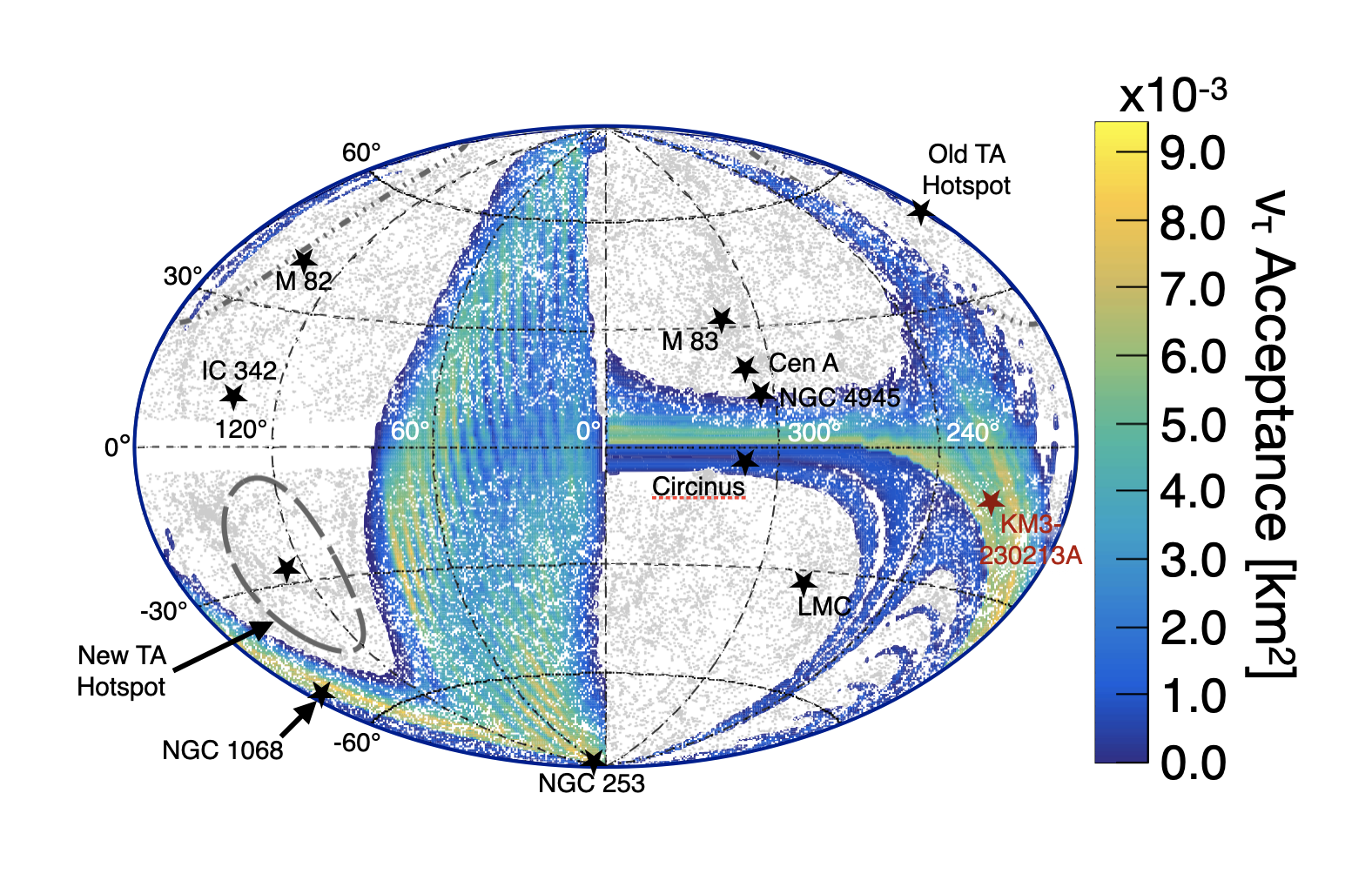}
    \hfil
    \begin{minipage}{\dimexpr 0.5\linewidth-\columnsep}
        \captionsetup{singlelinecheck=off, skip=0pt}
        \caption{PBR sky coverage in Galactic coordinates assuming a 20-day flight for a
          balloon trajectory that follows historical wind
          patterns. The launch date was set at April 6, 2027 (new
          moon). The acceptances were calculated at $10^9~{\rm
            GeV}$. For reference, the black stars indicate selected
          nearby astrophysics sources and reported hotspots from the
          Telescope Array (TA) experiment, whereas the red star indicates
          the arrival direction of KM3-230213A.}
    \label{fig:skycoverage} 
    \end{minipage}
    \vspace{-2ex}
\end{figure}
    
The energy-squared weighted flux of the transient neutrino source discussed in Sec.~\ref{sec:3} is 
\begin{equation}
  E_\nu^2 \ \frac{dN}{dE_\nu}  \sim 10^{-6}~{\rm GeV \ cm^{-2}\
    s^{-1}} \, ,
\end{equation}
see Fig.~1 in~\cite{Neronov:2025jfj}.  In rescaling $dN/dE_\nu$ by a factor of $100$ to approximately account for the combination of exposure and detector size of IceCube relative to KM3NeT~\cite{Anchordoqui:2025xug,Yuan:2025zwe}, we obtain
\begin{equation}
  \left. E_\nu^2 \ \frac{dN}{dE_\nu} \right|_{\rm BSM} \sim 10^{-4}~{\rm GeV \ cm^{-2} \
    s^{-1}} \, .
\end{equation}
For this flux of BSM messegers, we expect a fluence over an observation time
of $1000~{\rm s}$ to be ${\cal F}_{\rm BSM} \sim 0.1~{\rm GeV}\ {\rm cm}^{-2}$
in the energy range of $10^{7.8} \lesssim E_\nu/{\rm GeV} \lesssim
10^{9.0}$. For an observation time of 20 days, we expect the fluence
to be ${\cal F}_{\rm BSM} \sim 170~{\rm GeV} \ {\rm cm}^{-2}$ over
this energy range. Both fluences are within the
projected 90\%~CL sensitivity of PBR shown in
Fig.~\ref{fig:3}. Furthermore, as one can check by inspection of
Fig.~\ref{fig:skycoverage},
PBR's sky coverage (20 day mission with April 2027 launch) includes the KM3-230213A arrival
direction (and a longer flight would cover a larger fraction of the sky).

\vspace{-2ex}

\section{Conclusions}
\label{sec:5}

PBR is well-positioned to observe nearby bursting neutrino sources
with short duration (viz., emission lasting $\lesssim 1000~{\rm s}$). If KM3-230313A happened to originate from
such a nearby transient, with a possible example being the rapid
emission of energetic Hawking radiation from a PBH near the end of its evaporation lifetime, then PBR may be able to make a similar
observation. On the other hand,
KM3NeT and PBR have advantageous chord lengths in the arrival
direction of KM3-230213A. This will allow PBR to be sensitive to BSM
messengers (which produce $\tau$-leptons upon arrival to Earth) from sources at cosmological distances.

\section*{Acknowledgments}
\noindent
\small{ The authors would like to acknowledge the support by NASA
  awards RTOP 21-APRA21-0071, 80NSSC22K1488, and 80NSSC24K1780, by the French space agency
  CNES and the Italian Space agency ASI. The work is supported by OP
  JAC financed by ESIF and the MEYS
  CZ.02.01.01/00/22\_008/0004596. The work of L.A.A. and K.P.C. is
  supported by the US National Science Foundation (grant
  PHY-2412679). We gratefully acknowledge the collaboration and expert
  advice provided by the PUEO collaboration. We also acknowledge the
  invaluable contributions of the administrative and technical staffs
  at our home institutions.  }

%


\vspace{-2ex}

\begin{thebibliography}{99}

\setlength{\itemsep}{-0em}
\footnotesize

\bibitem{KM3Net:2016zxf}
KM3Net collaboration, J. Phys. G {\bf 43} (2016) 084001 [arXiv:1601.07459].

\bibitem{KM3NeT:2022pnv}
KM3NeT collaboration, JINST {\bf 17} (2022) P07038 [arXiv:2203.10048].

\bibitem{KM3NeT:2025npi}
KM3NeT collaboration, Nature {\bf 638} (2025) 376.

\bibitem{ICRC2025Eser}
J. Eser for the JEM-EUSO collaboration, PoS ICRC2025 (2025) 249. 

\bibitem{POEMMA:2020ykm}
POEMMA collaboration, JCAP {\bf 06} (2021) 007 [arXiv:2012.07945].


\bibitem{JEM-EUSO:2023ypf}
JEM-EUSO collaboration, Astropart. Phys. {\bf 154} (2024) 102891 [arXiv:2401.06525].




\bibitem{Adams:2025owi}
J.~H.~Adams \textit{et al.}, arXiv:2505.20762.


\bibitem{Battisti:2024rfv}
JEM-EUSO collaboration, Nucl. Instrum. Meth. A {\bf 1068} (2024) 169727 [arXiv:2408.14867].


\bibitem{Abarr_2021}
PUEO Collaboration, JINST {\bf 16} (2021) P08035.

\bibitem{Fedynitch:2018cbl}
A. Fedynitch, F. Riehn, R. Engel, T.K. Gaisser and T. Stanev, Phys. Rev. D {\bf 100} (2019) 103018 [arXiv:1806.04140].

\bibitem{Gaisser:2011klf}
T.~K.~Gaisser,
Astropart. Phys. {\bf 35} (2012) 801-806 
[arXiv:1111.6675].


\bibitem{Illana:2009qv}
J.I. Illana, M. Masip and D. Meloni, JCAP {\bf 09} (2009) 008 [arXiv:0907.1412].


\bibitem{Illana:2010gh}
J.I. Illana, P. Lipari, M. Masip and D. Meloni, Astropart. Phys. {\bf 34} (2011) 663 [arXiv:1010.5084].

\bibitem{Groom:2001kq}
D.E. Groom, N.V. Mokhov and S.I. Striganov, Atom. Data Nucl. Data Tabl. {\bf 78} (2001) 183.

\bibitem{mu-eloss-web}
Particle Data Group, 
{\tt https://pdg.lbl.gov/2023/AtomicNuclearProperties/HTML/water$_-$liquid.html}

\bibitem{Abramowicz:1997ms}
H. Abramowicz and A. Levy, arXiv:hep-ph/9712415.


\bibitem{Abramowicz:1991xz}
H. Abramowicz, E.M. Levin, A. Levy and U. Maor, Phys. Lett. B {\bf 269} (1991) 465.


\bibitem{Lohmann:1985qg}
W.~Lohmann, R.~Kopp and R.~Voss,
doi:10.5170/CERN-1985-003.

\bibitem{Lipari:1991ut}
P. Lipari and T. Stanev, Phys. Rev. D {\bf 44} (1991) 3543.


\bibitem{Dutta:2000hh}
S.I. Dutta, M.H. Reno, I. Sarcevic and D. Seckel, Phys. Rev. D {\bf 63} (2001) 094020 [arXiv:hep-ph/0012350].



\bibitem{PierreAuger:2023pjg}
Pierre Auger collaboration, PoS ICRC2023 (2023) 1488.


\bibitem{IceCube:2025ezc}
IceCube collaboration, arXiv:2502.01963.


\bibitem{Li:2025tqf}
S.W. Li, P. Machado, D. Naredo-Tuero and T. Schwemberger, arXiv:2502.04508.

\bibitem{KM3NeT:2025ccp}
KM3NeT collaboration, arXiv:2502.08173.


\bibitem{Neronov:2025jfj}
A. Neronov, F. Oikonomou and D. Semikoz, arXiv:2502.12986.

\bibitem{Yuan:2025zwe} 
C.~Yuan, L.~Pfeiffer, W.~Winter, S.~Buson, F.~Testagrossa, J.~M.~S.~Zaballa and A.~Azzollini, arXiv:2506.21111.

\bibitem{Airoldi:2025opo}
L.F.T. Airoldi, G.F.S. Alves, Y.F. Perez-Gonzalez, G.M. Salla and R.Z. Funchal, arXiv:2505.24666.

\bibitem{Fang:2025nzg}
K. Fang, F. Halzen and D. Hooper, Astrophys. J. Lett. {\bf 982} (2025) L16 [arXiv:2502.09545].


\bibitem{Crnogorcevic:2025vou}
M.~Crnogor{\v{c}}evi{\'c}, C. Blanco and T. Linden, arXiv:2503.16606.


\bibitem{Klipfel:2025jql} 
A.P. Klipfel and D.I. Kaiser, 
arXiv:2503.19227.

\bibitem{Anchordoqui:2025xug}
L.A. Anchordoqui, F. Halzen and D. L\"ust, arXiv:2505.23414.


\bibitem{Anchordoqui:2025opy}
L.A. Anchordoqui, A. Bedroya and D. L\"ust, arXiv:2506.14874.


\bibitem{Sakharov:2025oev}
A.~S.~Sakharov, R.~Konoplich and M.~Gogberashvili, arXiv:2506.23387.


\bibitem{Brdar:2025azm}
V. Brdar and D.S. Chattopadhyay, arXiv:2502.21299.


\bibitem{Farzan:2025ydi}  
Y. Farzan and M. Hostert, arXiv:2505.22711.


\bibitem{Dev:2025czz} 
P.S.B. Dev, B. Dutta, A. Karthikeyan, W. Maitra, L.E. Strigari and A. Verma, arXiv:2505.22754.

\bibitem{Reno:2019jtr}
M.~H.~Reno, J.~F.~Krizmanic and T.~M.~Venters,
Phys. Rev. D 100 (2019) 063010
[arXiv:1902.11287].

\bibitem{Venters:2019xwi}
T.M. Venters, M.H. Reno, J.F. Krizmanic, L.A. Anchordoqui, C. Gu\'epin and A.V. Olinto, Phys. Rev. D {\bf 102} (2020) 123013 [arXiv:1906.07209].

\bibitem{Anchordoqui:2002vb}
L.~A.~Anchordoqui, J.~L.~Feng, H.~Goldberg and A.~D.~Shapere,
Phys. Rev. D \textbf{66} (2002) 103002. 

\bibitem{Feldman:1997qc}
G.~J.~Feldman and R.~D.~Cousins,
Phys. Rev. D \textbf{57}, 3873-3889 (1998)
[arXiv:physics/9711021].

\bibitem{ANTARES:2017bia}
A. Albert et al., Astrophys. J. {\bf 850} (2017) L35 [arXiv:1710.05839].







\end{thebibliography}
\end{document}